\documentclass[sigconf]{acmart}
\AtBeginDocument{%
  }

\setcopyright{acmlicensed}
\copyrightyear{2018}
\acmYear{2018}
\acmDOI{XXXXXXX.XXXXXXX}
\acmConference[Conference acronym 'XX]{Make sure to enter the correct
  conference title from your rights confirmation email}{June 03--05,
  2018}{Woodstock, NY}
\acmISBN{978-1-4503-XXXX-X/2018/06}

\usepackage{xcolor}
\usepackage{multirow}
\begin{document}

\title{Text-Guided Visual Representation Learning for Robust Multimodal E-Commerce Recommendation} 

\author{Yufei Guo}
\email{guoyufei21@mails.tsinghua.edu.cn}
\affiliation{%
  \institution{Tsinghua University}
  \state{Beijing}
  \country{China}}

\author{Jing Ma}
\email{ma-j22@mails.tsinghua.edu.cn}
\affiliation{%
  \institution{Tsinghua University}
  \state{Beijing}
  \country{China}}

\author{Tianlu Zhang}
\email{tlzhang@mail.tsinghua.edu.cn}
\affiliation{%
  \institution{Tsinghua University}
  \state{Beijing}
  \country{China}}

\author{Shijie Yang}
\email{yangshijie30@jd.com}
\affiliation{%
  \institution{JD.COM}
  \state{Beijing}
  \country{China}}

\author{Yanlong Zang}
\email{zangyanlong1@jd.com}
\affiliation{%
  \institution{JD.COM}
  \state{Beijing}
  \country{China}}

\author{Weijie Ding}
\email{dingweijie8@jd.com}
\affiliation{%
  \institution{JD.COM}
  \state{Beijing}
  \country{China}}

\author{Pinghua Gong}
\email{gongpinghua1@jd.com}
\affiliation{%
  \institution{JD.COM}
  \state{Beijing}
  \country{China}}

\author{Jungong Han}
\email{jghan@mail.tsinghua.edu.cn}
\affiliation{%
  \institution{Tsinghua University}
  \state{Beijing}
  \country{China}}

\renewcommand{\shortauthors}{Trovato et al.}

\begin{abstract}
Multimodal item embeddings are crucial for e-commerce item-to-item (I2I) retrieval, yet real-world product images often contain promotional overlays and background clutter that inject spurious visual cues and degrade retrieval robustness. This issue is particularly pronounced in MLRM-style pipelines, where a frozen vision encoder is connected to an LLM through a lightweight connector that must selectively aggregate visual tokens.
We propose Text-Guided Q-Former (TGQ-Former), a text-guided visual representation learning framework that leverages structured metadata as semantic guidance for visual token extraction while preserving complementary visual evidence. Concretely, TGQ-Former employs a hybrid-query connector to disentangle metadata-anchored and exploratory visual streams, and introduces a lightweight reliability-aware dual-gated vector modulation module to adaptively calibrate their contributions under noisy inputs.
Experiments on large-scale, real-world e-commerce datasets with full-pool retrieval show that TGQ-Former consistently outperforms strong connector baselines and end-to-end MLLMs. On average, it improves Hit Rate@100 (H@100) by 6.04\%, demonstrating the effectiveness of text-guided visual encoding for robust multimodal retrieval.
\end{abstract}




\maketitle

\section{Introduction}
Item-to-item (I2I) retrieval is a core component of large-scale e-commerce recommendation systems.
It typically operates as a high-recall candidate generation stage, which requires learning retrieval-oriented and discriminative item embeddings under strict latency and throughput constraints~\cite{Covington2016DeepNN}.
In modern product catalogs, items are typically associated with both visual and structured textual metadata (e.g., title, brand, and hierarchical categories): text offers structured semantic signals, while images capture fine-grained visual attributes.
In light of this, the core objective of our work is to learn high-quality multimodal item embeddings, which enable retrieval performance that surpasses single-modality models~\cite{Wei2019MMGCNMG, Chen2019PersonalizedFR}.

\begin{figure}[!t]
\centering
\includegraphics[width=0.8\linewidth]{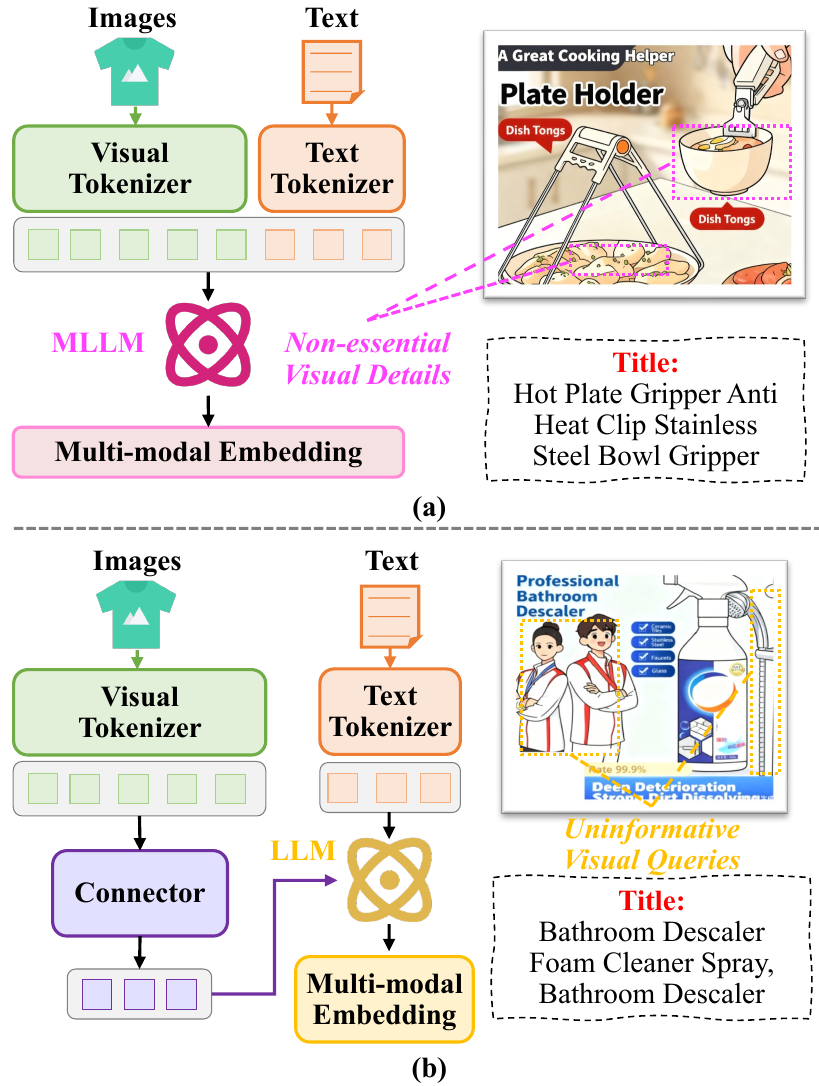}
\caption{Illustration of different strategies for using LLMs to enhance multimodal representations. (a) MLLMs based methods. (b) MLRMs based methods.}
\label{pic:intro}
\vspace{-4mm}
\end{figure}

In real-world e-commerce, the text modality encompasses product titles and manually curated hierarchical titles at all levels, which can deliver precise semantic information about products.
By contrast, product images uploaded by merchants are rarely clean, object-centric photographs.
Many of them come from \emph{poster-style listings}, where the product is overlaid with various promotional elements (e.g., sale text, price tags, coupons) and decorative layouts.
These non-product visual signals can have mixed effects on item representation:
certain contextual presentation details (e.g., standard product packaging, co-displayed matching accessories) can help distinguish visually similar items and even convey their potential application scenarios; 
in contrast, promotional overlays introduce systematic, marketing-driven visual artifacts that merely correlate with campaign strategies rather than the intrinsic similarity of the products themselves.



Recent advances in LLMs and MLLMs have motivated their adoption in e-commerce recommendation~\cite{Wu2023ASO,Hou2023LargeLM,Xu2025ASO}.
One common paradigm is to directly use an MLLM to map images and text into a unified embedding (Fig.~\ref{pic:intro}(a)), but it is often inefficient for industrial item-to-item (I2I) retrieval.
Moreover, MLLMs are typically optimized for generic visual understanding, and their generated representations tend to encode spurious artifacts in e-commerce imagery (e.g., promotional overlays and decorative layouts) that have only weak correlations with intrinsic item similarity. 
Alternatively, \emph{Multimodal Large Representation Models (MLRMs)} connect a frozen vision encoder to an LLM via a lightweight connector (Fig.~\ref{pic:intro}(b))~\cite{zhang2025notellm}, offering improved modularity and deployment flexibility.
However, existing MLRM approaches overlook the aforementioned image quality issues in e-commerce imagery, which consequently causes their connectors to aggregate visual tokens using learnable queries or naive cross-attention that lack domain-specific semantic grounding.
Attention is thus prone to being dominated by visually salient yet product-irrelevant regions, ultimately yielding unstable or distorted item representations in poster-style product listings.

\begin{figure}[!t]
\centering
\includegraphics[width=0.85\linewidth]{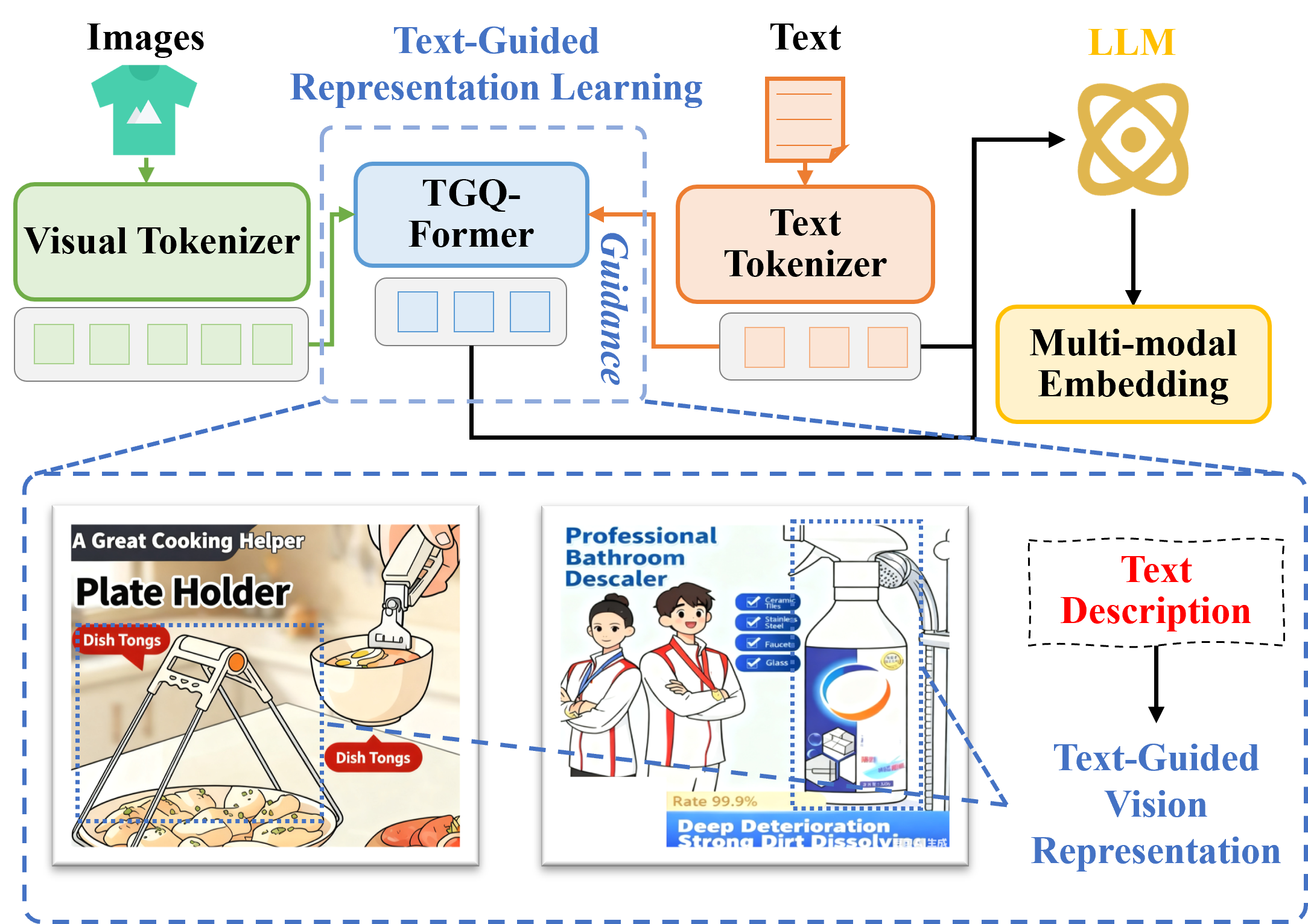}
\caption{Illustration of our proposed text-guided visual representation learning paradigm.}
\label{pic:intro2}
\vspace{-4mm}
\end{figure}

Moreover, we find that simple, training-free preprocessing heuristics (e.g., center cropping) offer only marginal improvements in the presence of promotional overlays (Appendix~\ref{app:crop_baselines}).
This observation indicates that indiscriminate suppression of non-product regions is impractical in real-world scenarios: valuable contextual cues often coexist with detrimental overlay artifacts, making blind exclusion of non-product content counterproductive. 
To robustly leverage images in poster-style listings and meet the demands of industrial e-commerce I2I retrieval, a selective approach to utilizing visual evidence is required, rather than a one-size-fits-all multimodal fusion strategy.
On one hand, structured metadata (e.g., hierarchical categories and brand) serves as a stable semantic anchor, enabling the model to prioritize product-relevant visual cues and alleviate the adverse impact of overlay artifacts.
On the other hand, product images inherently contain complementary visual signals that are not fully specified by text and exhibit item-wise variability, including fine-grained appearance details and useful presentation context in real e-commerce listings.
These observations motivate a text-guided visual representation learning framework, referred to as \textbf{Text-Guided Q-Former (TGQ-Former)}, that uses metadata to steer visual extraction while retaining exploratory learnable queries to capture complementary evidence beyond explicit text.


Specifically, we first propose the \textbf{Hybrid-Query Connector}, a connector integrated with a hybrid query set that balances the trade-off between \emph{metadata-anchored} and \emph{exploratory} visual extraction. 
Metadata-anchored semantic queries leverage structured metadata as a semantic prior to focus on product-relevant visual regions, while exploratory metadata-agnostic learnable queries capture complementary visual signals that lie beyond the scope of explicit textual metadata.
Second, we propose the \textbf{Dual-Gated Vector Modulation}, a lightweight feature modulation mechanism that adaptively calibrates the two query streams prior to their input into the LLM. Specifically, we modulate the gating of the metadata-anchored query stream according to an image-metadata semantic agreement signal, and regulate the gating of the exploratory query stream based on an item-specific visual reliability signal. 
Finally, we design the \textbf{Redundancy-Reduction Regularizer} to enforce the two query streams to be mutually complementary, and optimize a contrastive retrieval objective by leveraging item embeddings generated by the LLM.

Our contributions are summarized as follows:
\begin{itemize}
  \item We identify and characterize \emph{poster-style listings} in real-world e-commerce imagery as a systematic source of spurious visual artifacts,
        where representation learning can drift toward promotional overlay patterns rather than product identity, degrading fine-grained I2I retrieval.
  \item We propose \textbf{TGQ-Former}, a text-guided visual representation learning framework for I2I retrieval.
        At its core, TGQ-Former introduces a \textbf{Hybrid-Query Connector} that disentangles visual evidence into two query streams---metadata-anchored semantic queries and exploratory learnable queries---enabling selective extraction of product-relevant and complementary cues with structured metadata.
  \item We introduce \textbf{Dual-Gated Vector Modulation} together with a \textbf{Redundancy-Reduction Regularizer} to perform reliability-aware calibration and complementarity enforcement between the two streams, improving robustness under poster-style and otherwise noisy listings.
\end{itemize}

\section{Related Work}

\subsection{Vision--Language Pretraining}
Large-scale vision--language pretraining has become a dominant paradigm for multimodal representation learning.
Early works such as UNITER~\cite{Chen2019UNITERLU}, OSCAR~\cite{Li2020OscarOA}, and ALIGN~\cite{Jia2021ScalingUV} demonstrated that aligning image and text embeddings enables strong transfer to downstream tasks.
Contrastive methods like CLIP~\cite{Radford2021LearningTV} and Florence~\cite{Yuan2021FlorenceAN} further scale this paradigm to web-scale datasets, achieving remarkable zero-shot generalization.
More recent models such as BLIP~\cite{Li2022BLIPBL} and BLIP-2~\cite{li2023blip} integrate generative and contrastive objectives, and leverage a Query Transformer (Q-Former) to bridge frozen image encoders with LLM backbones.
While highly effective for open-domain vision--language tasks, these models are typically trained on data distributions that differ from real-world e-commerce imagery, where promotional overlay artifacts and clutter are prevalent.
Our work builds on this line by adapting query-based connectors to retrieval-oriented I2I recommendation, with an emphasis on reliability-aware visual evidence extraction in poster-style listings.

\subsection{Query-Based Architectures and Robust Cross-Modal Extraction}
Query-based architectures have shown promise for efficient cross-modal token extraction.
Perceiver~\cite{Jaegle2021PerceiverGP} and Perceiver-IO~\cite{Jaegle2021PerceiverIA} demonstrated that a fixed set of latent queries can aggregate information from high-dimensional inputs.
Similarly, BLIP-2's Q-Former~\cite{li2023blip} uses trainable queries to extract visual features aligned with language backbones.
In noisy e-commerce settings, however, relying on a single, semantically unstructured query set can cause cross-attention to be dominated by visually salient yet product-irrelevant regions (e.g., promotional overlay artifacts and decorative layouts), yielding biased representations.
Prior work in vision--language grounding and robust representation learning suggests that incorporating semantic priors and confidence/quality signals can improve robustness by steering attention toward task-relevant evidence under diverse input conditions (e.g., grounding with textual/category cues or object-centric priors)~\cite{Zhang2021VinVLMV, Desai2020VirTexLV}.
In contrast to methods that inject only a single form of prior, our approach disentangles \emph{anchored semantic evidence} and \emph{exploratory visual evidence} with a hybrid query set, and further performs reliability-aware calibration to prevent overlay artifacts in poster-style listings from dominating retrieval representations.

\subsection{Item-to-Item Recommendation}
Item-to-item (I2I) recommendation is a fundamental component of large-scale recommender systems, often instantiated as a high-recall candidate generation stage prior to ranking~\cite{Covington2016DeepNN, Johnson2017BillionScaleSS}.
Traditional methods rely on collaborative filtering or shallow content features, whereas recent advances incorporate multimodal signals to enhance item representations.
For example, MAPS~\cite{Das2022MAPSMA}, ECLIP~\cite{Jin2023LearningIR}, and FAME-VLM~\cite{Han2023FAMEViLMV} exploit product text and images with relatively lightweight, task-specific encoders to generate embeddings.
NoteLLM-2~\cite{zhang2025notellm} further explores LLM-enhanced multimodal embeddings for retrieval-style I2I recommendation, highlighting the potential of LLMs in industrial retrieval pipelines.
However, existing approaches can remain vulnerable to real-world image noise and often lack mechanisms to selectively utilize visual information when it becomes unreliable, leading to brittle nearest-neighbor retrieval in poster-style listings.
Our method complements this line by introducing a two-stream, query-based connector with reliability-aware modulation and complementarity regularization, improving robustness and fine-grained retrieval quality in large-scale e-commerce I2I recommendation.

\section{Preliminary}

\subsection{Problem Formulation}
In item-to-item (I2I) retrieval, each item corresponds to a product with a primary image and associated textual metadata.
We consider a product set $\mathcal{I}=\{n_1,n_2,\ldots,n_m\}$, where each product is represented by $n_i=(v_i,x_i)$.
Here $v_i$ is the primary product image, and $x_i$ denotes the textual metadata, which may include a free-form title $t_i$ and structured fields such as brand $b_i$ and hierarchical categories $\mathbf{c}_i=(c_i^{(1)},c_i^{(2)},c_i^{(3)})$.

Given a query product $n_i$, the goal is to retrieve a ranked list of relevant items from $\mathcal{I}$.
We learn an embedding function $f(\cdot)$ that maps each product to a vector representation $\mathbf{z}_i=f(v_i,x_i)\in\mathbb{R}^d$.
The relevance between two products is computed by a similarity function (e.g., cosine similarity) $\text{sim}(\mathbf{z}_i,\mathbf{z}_j)$,
and the system returns items with the highest similarity to the query.

\begin{figure*}[!t]
\centering
\includegraphics[width=0.85\linewidth]{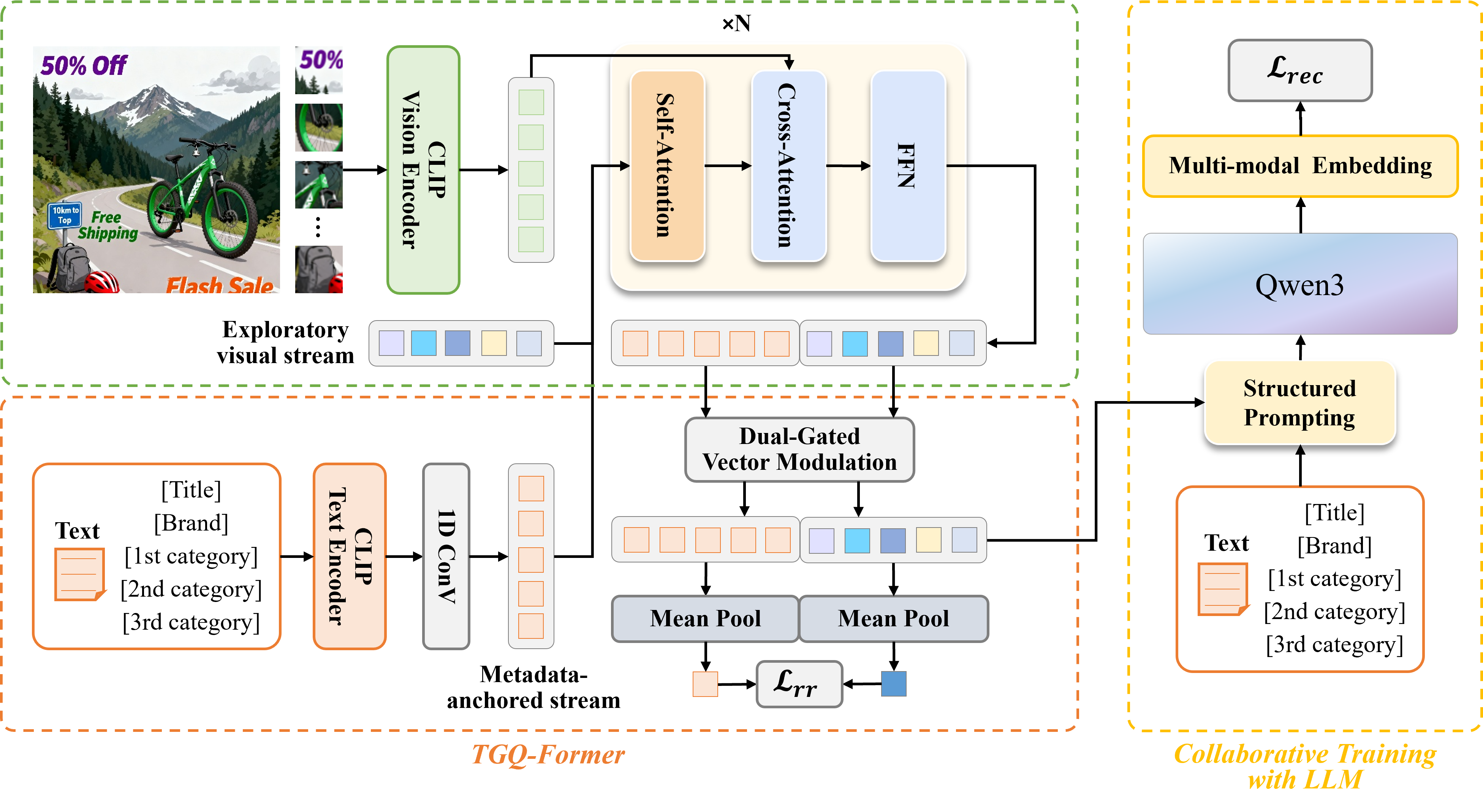}
\caption{Architecture of the proposed Text-Guided Visual Representation Learning framework.}
\label{pic:TGQ-Former}
\end{figure*}

\subsection{Basic MLRM Pipeline}
We follow the standard MLRM pipeline in Fig.~\ref{pic:intro}(b).
Given an item $n_i=(v_i,x_i)$, we encode the image $v_i$ with a frozen vision encoder $V_{\theta}$ (e.g., CLIP) to obtain visual tokens
$\mathbf{H}_{img,i} \in \mathbb{R}^{L_v \times d_v}$, optionally along with a global image embedding $\mathbf{f}_{img,i}^{global} \in \mathbb{R}^{d_v}$.
A lightweight connector $C_{\pi}$ then extracts a compact set of visual embeddings in the LLM embedding space,
$\mathbf{E}^{(v)}_i \in \mathbb{R}^{L_c \times d_{llm}}$, where $d_{llm}$ is the LLM hidden size.
A common instantiation of $C_{\pi}$ is a query-based connector (e.g., Q-Former~\cite{li2023blip}), which uses $L_c$ learnable queries to cross-attend to $\mathbf{H}_{img,i}$ and summarize the visual content into $\mathbf{E}^{(v)}_i$.

\textbf{Prompt-based multimodal injection.}
To obtain a single-vector representation suitable for high-throughput retrieval, we inject $\mathbf{E}^{(v)}_i$ into an LLM prompt.
Concretely, we construct a textual prompt from metadata fields in $x_i$ and reserve a special placeholder token $\texttt{<IMG>}$ for visual embeddings:
\begin{center}
\fcolorbox{black}{gray!10}{\parbox{.9\linewidth}{
\textbf{Prompt}: Product image: \{`image': \texttt{<IMG>}\}, Product metadata: \{`brand': $b_i$, `category': $\mathbf{c}_i$, `title': $t_i$\}. Produce a single embedding for retrieval.
}}
\end{center}
After the LLM token embedding layer, the prompt yields text embeddings $\mathbf{E}^{(t)}_i \in \mathbb{R}^{T \times d_{llm}}$.
We replace the placeholder token at the predefined $\texttt{<IMG>}$ position with the visual embedding sequence $\mathbf{E}^{(v)}_i$,
resulting in a fused multimodal sequence $\mathbf{E}^{(m)}_i \in \mathbb{R}^{(T-1+L_c)\times d_{llm}}$.
The LLM processes $\mathbf{E}^{(m)}_i$ to produce hidden states,
and we take the hidden state at a designated summary position (e.g., the last token) as the final item representation $\mathbf{z}_i$.

\textbf{Instantiation in this work.}
In TGQ-Former, we instantiate $C_{\pi}$ as a \emph{Hybrid-Query Connector (HQC)} that produces two complementary visual embedding streams, i.e., metadata-anchored stream and exploratory stream.
A \emph{Dual-Gated Vector Modulation} module then calibrates the two streams before they are injected into the LLM through the $\texttt{<IMG>}$ placeholder (Fig.~\ref{pic:TGQ-Former}).

\section{Method}
\label{sec:method}

We study item-to-item (I2I) retrieval in e-commerce, where each item $n_i=(v_i,x_i)$ is associated with an image $v_i$ and textual metadata $x_i$ (e.g., title, brand, and hierarchical categories).
A practical challenge is that images from poster-style listings often contain promotional overlay artifacts and background clutter, which can introduce spurious cues if visual evidence is fused indiscriminately.
Therefore, our key design principle is \emph{adaptive disentanglement}: we extract two complementary streams of visual evidence and calibrate their contributions based on reliability.

Concretely, we disentangle visual evidence into (i) \emph{metadata-anchored semantic evidence} guided by structured metadata, which helps focus on product-relevant regions and suppress overlay artifacts, and (ii) \emph{exploratory visual evidence} that captures complementary appearance and presentation cues beyond explicit textual metadata.
We then \emph{adaptively modulate} the two streams using image–metadata agreement and visual reliability signals.

TGQ-Former consists of three components:
(1) a \textbf{Hybrid-Query Connector} that produces two complementary sets of LLM-aligned visual tokens (metadata-anchored vs.\ exploratory);
(2) \textbf{Dual-Gated Vector Modulation} that performs reliability-aware, channel-wise calibration using a centered-sigmoid gate; and
(3) a \textbf{Redundancy-Reduction Regularizer} to encourage complementarity.

\subsection{Hybrid-Query Connector}
\label{sec:tgq}

TGQ-Former extracts retrieval-oriented visual evidence from noisy e-commerce images by combining a metadata-anchored stream and an exploratory stream.
Given an item $n_i=(v_i,x_i)$ with image $v_i$ and textual metadata $x_i$, we use a frozen CLIP vision encoder to obtain visual tokens and a global image embedding. For clarity, we omit the item index $i$ when the context is unambiguous:
\begin{equation}
\mathbf{H}_{img}=\text{CLIP}_{vis}(v)\in\mathbb{R}^{L_v\times d_v},\qquad
\mathbf{f}^{global}_{img}=\text{Pool}(\mathbf{H}_{img})\in\mathbb{R}^{d_v},
\end{equation}
where $\text{Pool}(\cdot)$ denotes the encoder's global pooling (e.g., CLS token or average pooling).
We also encode textual metadata with a frozen CLIP text encoder to obtain token embeddings:
\begin{equation}
\mathbf{H}_{txt}=\text{CLIP}_{txt}(x)\in\mathbb{R}^{L_t\times d_v}.
\end{equation}

\textbf{Metadata-guided semantic query initialization.}
Unlike BLIP-2's vanilla Q-Former that uses learnable queries without domain-specific priors, we initialize a subset of queries from item metadata to provide semantic anchors for I2I retrieval.
Concretely, we compress $\mathbf{H}_{txt}$ into a fixed number of query slots using a 1D convolutional downsampling layer:
\begin{equation}
\tilde{\mathbf{H}}_{txt}=\text{Conv1D}(\mathbf{H}_{txt}; k, s)\in\mathbb{R}^{T_g\times d_v},
\end{equation}
where $k$ is the kernel size and $s$ is the stride. We set $T_g=\left\lceil\frac{L_t}{s}\right\rceil$ (after padding if needed), yielding a deterministic number of semantic query slots.
We then project the compressed embeddings into the query space:
\begin{equation}
\mathbf{Q}_{txt}=\tilde{\mathbf{H}}_{txt}\mathbf{W}_Q\in\mathbb{R}^{T_g\times d_q},
\end{equation}
where $\mathbf{W}_Q\in\mathbb{R}^{d_v\times d_q}$ is learnable. The resulting $\mathbf{Q}_{txt}$ provides metadata-anchored semantic priors for steering cross-attention toward product-relevant regions in poster-style listings.

\textbf{Learnable exploratory queries.}
To capture complementary visual evidence beyond explicit textual metadata (e.g., fine-grained appearance and presentation cues), we introduce an additional set of learnable queries
$\mathbf{Q}_{rnd}\in\mathbb{R}^{T_r\times d_q}$ initialized randomly.

\textbf{Hybrid-Query Connector.}
We concatenate the two query sets as the connector input,
$\mathbf{Q}_{in}=[\mathbf{Q}_{txt};\mathbf{Q}_{rnd}]\in\mathbb{R}^{(T_g+T_r)\times d_q}$,
and feed them into a Q-Former-style \textbf{Hybrid-Query Connector} that applies self-attention over queries and cross-attention over visual tokens:
\begin{equation}
\mathbf{E}^{(v)} = [\mathbf{E}_v^{txt};\mathbf{E}_v^{rnd}]
=\text{HQC}(\mathbf{Q}_{in},\mathbf{H}_{img}),
\end{equation}
where $\mathbf{E}_v^{txt}\in\mathbb{R}^{T_g\times d_{\text{llm}}}$ denotes metadata-anchored visual embeddings and
$\mathbf{E}_v^{rnd}\in\mathbb{R}^{T_r\times d_{\text{llm}}}$ denotes exploratory visual embeddings aligned to the downstream LLM hidden size $d_{\text{llm}}$.

\subsection{Dual-Gated Vector Modulation}
\label{sec:dual_gate}

Images from poster-style listings exhibit diverse noise patterns: promotional overlay artifacts and background clutter can dominate visually salient regions, and the usefulness of metadata anchoring can vary across items due to cross-modal disagreement.
To make the two evidence streams robust in such conditions, we introduce a dual-gated, channel-wise modulation module that calibrates the metadata-anchored stream and the exploratory stream \emph{separately}.
Unlike scalar gating, our gates output feature-wise vectors to enable \emph{partial trust}: unreliable channels can be suppressed while informative channels are preserved.

\textbf{Agreement signal from product titles.}
We estimate cross-modal agreement using the product title text, which provides concise, item-specific semantics.
Let $\mathcal{T}^{title}$ denote the title of the current item.
We obtain a global title embedding with the frozen CLIP text encoder:
\begin{equation}
\mathbf{f}^{global}_{title}=\text{Pool}\big(\text{CLIP}_{txt}(\mathcal{T}^{title})\big)\in\mathbb{R}^{d_v},
\end{equation}
and define the image--title agreement score as
\begin{equation}
S_{title}=\cos(\mathbf{f}^{global}_{img},\mathbf{f}^{global}_{title})\in[-1,1].
\end{equation}

\textbf{Stream summaries.}
Given TGQ-Former outputs $\mathbf{E}_v^{txt}\in\mathbb{R}^{T_g\times d_{llm}}$ and $\mathbf{E}_v^{rnd}\in\mathbb{R}^{T_r\times d_{llm}}$,
we summarize each stream by mean pooling:
\begin{equation}
\bar{\mathbf{e}}_{txt}=\text{Mean}(\mathbf{E}_v^{txt})\in\mathbb{R}^{d_{llm}},\quad
\bar{\mathbf{e}}_{rnd}=\text{Mean}(\mathbf{E}_v^{rnd})\in\mathbb{R}^{d_{llm}}.
\end{equation}

\textbf{Vector gate parameterization.}
We parameterize channel-wise gates with a centered sigmoid,
\begin{equation}
\text{cSigmoid}(\mathbf{x})=2\sigma(\mathbf{x})-1\in(-1,1),
\end{equation}
and denote the resulting gate vector as $\boldsymbol{\beta}\in(-1,1)^{d_{llm}}$.
We apply residual rescaling $(1+\boldsymbol{\beta})\in(0,2)^{d_{llm}}$ to each token embedding (broadcast over the token dimension), enabling smooth attenuation or amplification per channel.

\textbf{Gating networks.}
Both gates are implemented as lightweight two-layer MLPs that map their inputs to a $d_{llm}$-dimensional vector:
\begin{equation}
\mathcal{G}_{\ast}(\mathbf{u})=\mathbf{W}_2\,\phi(\mathbf{W}_1\mathbf{u})\in\mathbb{R}^{d_{llm}},
\end{equation}
where $\ast\in\{txt,rnd\}$, $\phi(\cdot)$ is a non-linear activation (e.g., GELU), and $\mathbf{W}_1,\mathbf{W}_2$ are learnable parameters.
We use separate parameter sets for $\mathcal{G}_{txt}$ and $\mathcal{G}_{rnd}$.

\textbf{Gate for metadata-anchored semantics.}
The metadata-anchored stream should be emphasized when the image agrees with textual metadata and down-weighted when the agreement is low.
We therefore condition its gate on both the agreement signal and cross-stream context:
\begin{equation}
\mathbf{u}_{txt}=
[\mathbf{f}^{global}_{img};\mathbf{f}^{global}_{title};S_{title};\bar{\mathbf{e}}_{txt};\bar{\mathbf{e}}_{rnd}],
\end{equation}
\begin{equation}
\boldsymbol{\beta}_{txt}=\text{cSigmoid}\big(\mathcal{G}_{txt}(\mathbf{u}_{txt})\big)\in(-1,1)^{d_{llm}},
\end{equation}
\begin{equation}
\tilde{\mathbf{E}}_v^{txt}=\mathbf{E}_v^{txt}\odot(1+\boldsymbol{\beta}_{txt}).
\end{equation}

\textbf{Gate for exploratory evidence.}
The exploratory stream is beneficial for capturing complementary appearance cues, but it can also propagate spurious patterns under severe overlay artifacts in poster-style listings.
We thus condition its gate on image-derived and stream-specific signals:
\begin{equation}
\mathbf{u}_{rnd}=[\mathbf{f}^{global}_{img};\bar{\mathbf{e}}_{rnd}],
\end{equation}
\begin{equation}
\boldsymbol{\beta}_{rnd}=\text{cSigmoid}\big(\mathcal{G}_{rnd}(\mathbf{u}_{rnd})\big)\in(-1,1)^{d_{llm}},
\end{equation}
\begin{equation}
\tilde{\mathbf{E}}_v^{rnd}=\mathbf{E}_v^{rnd}\odot(1+\boldsymbol{\beta}_{rnd}).
\end{equation}

Finally, we concatenate the calibrated embeddings and inject them into the downstream LLM:
\begin{equation}
\mathbf{E}^{(v)}=[\tilde{\mathbf{E}}_v^{txt};\tilde{\mathbf{E}}_v^{rnd}]
\in\mathbb{R}^{(T_g+T_r)\times d_{llm}}.
\end{equation}

\subsection{Redundancy Reduction for Complementarity}
\label{sec:rr}

To encourage the metadata-anchored stream and the exploratory stream to encode complementary evidence, we adopt a redundancy reduction objective over each mini-batch.
For each item $i$ in a batch of size $B$, we pool the calibrated embeddings into stream-level vectors:
\begin{equation}
\mathbf{s}^{txt}_i=\text{Mean}(\tilde{\mathbf{E}}_{v,i}^{txt})\in\mathbb{R}^{d_{llm}},\quad
\mathbf{s}^{rnd}_i=\text{Mean}(\tilde{\mathbf{E}}_{v,i}^{rnd})\in\mathbb{R}^{d_{llm}}.
\end{equation}
Stacking them across the batch yields matrices $\mathbf{S}^{txt}=[\mathbf{s}^{txt}_1;\ldots;\mathbf{s}^{txt}_B]\in\mathbb{R}^{B\times d_{llm}}$ and
$\mathbf{S}^{rnd}=[\mathbf{s}^{rnd}_1;\ldots;\mathbf{s}^{rnd}_B]\in\mathbb{R}^{B\times d_{llm}}$.

We standardize each feature dimension across the mini-batch:
\begin{equation}
\hat{\mathbf{S}}^{txt}=\frac{\mathbf{S}^{txt}-\text{Mean}_B(\mathbf{S}^{txt})}{\text{Std}_B(\mathbf{S}^{txt})+\epsilon},\quad
\hat{\mathbf{S}}^{rnd}=\frac{\mathbf{S}^{rnd}-\text{Mean}_B(\mathbf{S}^{rnd})}{\text{Std}_B(\mathbf{S}^{rnd})+\epsilon},
\end{equation}
where $\text{Mean}_B(\cdot)$ and $\text{Std}_B(\cdot)$ are computed per feature dimension over the batch.

We then compute the cross-correlation matrix between the two streams:
\begin{equation}
\mathbf{C}=\frac{1}{B}(\hat{\mathbf{S}}^{txt})^\top\hat{\mathbf{S}}^{rnd}\in\mathbb{R}^{d_{llm}\times d_{llm}}.
\end{equation}
The redundancy reduction loss penalizes same-dimension cross-stream correlations:
\begin{equation}
\mathcal{L}_{rr}=\frac{1}{d_{llm}}\|\mathrm{diag}(\mathbf{C})\|_2^2
=\frac{1}{d_{llm}}\sum_{k=1}^{d_{llm}} C_{kk}^2,
\end{equation}
which discourages the two streams from encoding redundant evidence in the same subspace while allowing useful cross-dimension interactions.

\subsection{Joint Contrastive Training for I2I Retrieval}
\label{sec:training}

Following the MLRM prompting interface, we inject the calibrated visual embeddings
$\mathbf{E}^{(v)}=[\tilde{\mathbf{E}}_v^{txt};\tilde{\mathbf{E}}_v^{rnd}]$
into a pretrained LLM together with the metadata prompt (by replacing the $\texttt{<IMG>}$ placeholder), and obtain an item representation $\mathbf{z}_i$ from the LLM hidden states (e.g., the last-token hidden state).
We train the model for retrieval-style I2I recommendation with an InfoNCE-style contrastive objective over positive item pairs $(i,j)$ and in-batch negatives:
\begin{equation}
\mathcal{L}_{rec}= -\log
\frac{\exp(\text{sim}(\mathbf{z}_i,\mathbf{z}_j)/\tau)}
{\sum_{k\in \mathcal{B}\setminus\{i\}}\exp(\text{sim}(\mathbf{z}_i,\mathbf{z}_k)/\tau)},
\end{equation}
where $\mathcal{B}$ denotes the current mini-batch, $\text{sim}(\cdot,\cdot)$ is cosine similarity, and $\tau$ is the temperature.

The overall training objective is
\begin{equation}
\mathcal{L}=\mathcal{L}_{rec}+\lambda_{rr}\mathcal{L}_{rr}.
\end{equation}

\section{Experiments}
To systematically evaluate the effectiveness of our proposed framework, we conduct comprehensive experiments designed to answer the following research questions (RQs):

\begin{itemize}
\item \textbf{RQ1:} How does our overall framework perform compared to state-of-the-art multimodal recommendation methods and existing connector designs?
\item \textbf{RQ2:} How does our TGQ-Former-based MLRM paradigm compare with directly applying end-to-end MLLMs in terms of both retrieval quality and computational efficiency?
\item \textbf{RQ3:} What are the individual contributions of the Hybrid-Query Connector, Dual-Gated Vector Modulation, and the Redundancy-Reduction Regularizer?
\item \textbf{RQ4:} How do metadata-anchored and exploratory queries allocate cross-attention over visual regions in poster-style listings?
\end{itemize}

\subsection{Experiments Setup}

\subsubsection{\textbf{Datasets}}

We construct our dataset from one year of user ``add-to-cart'' behavior logs sourced from the Checkout Page Add-on channel of our e-commerce platform. To generate candidate item pairs, we employ the Swing algorithm~\cite{yang2020large} to recall the top-100 similar items for each trigger item.
Note that Swing is only used for constructing training/testing pairs; during evaluation, we perform full-pool retrieval over all test items (about 134K candidates) for every query.
 The final dataset contains approximately 3.5 million item pairs for training and 67,000 for testing. To prevent data leakage, the trigger items in the training and testing sets are strictly disjoint, with both sets exhibiting similar data distributions. Detailed dataset statistics are provided in Table~\ref{tab:dataset}.
\begin {table}[h]
\small
\centering
\setlength{\tabcolsep}{4pt}
\caption{Detailed statistics of training and testing datasets.}
\label{tab:dataset}
\begin{tabular}{l r l r}
\toprule
\multicolumn{4}{c}{\textbf{Training Dataset}} \\
\midrule
item pairs & 3,521,865 & unique items & 148,270 \\
avg. words per sku title & 46.85 & 1st categories & 54 \\
avg. words per sku brand & 5.66 & 2nd categories & 466  \\
 & & 3rd categories & 2820  \\
\toprule
\multicolumn{4}{c}{\textbf{Testing Dataset}} \\
\midrule
item pairs & 67,341 & unique items & 134,682 \\
avg. words per sku title & 48.02 & 1st categories & 51 \\
avg. words per sku brand & 5.40 & 2nd categories & 419 \\
 & & 3rd categories & 2096\\
\bottomrule
\end{tabular}
\end{table}
\begin{table*}
  \caption{Comparison of modality fusion methods across three different LLMs.}
  \label{tab:overall_performance}
  \resizebox{0.85\linewidth}{!}{
  \renewcommand{\arraystretch}{0.8}
  \small
  \begin{tabular}{c|c|cccccc}
    \toprule
    \textbf{Base Model} & \textbf{Connector} & H@1 & H@5 & H@10 & H@20 & H@50 & H@100 \\
    \midrule
    \multirow{7}{*}{\begin{tabular}{@{}c@{}}Chinese-CLIP-ViT-B/16 \\ + \\ Qwen3-Embedding-0.6B\end{tabular}} &  {LLaVA:linear} & $7.86$ & $20.16$ & $27.73$ & $37.18$ & $51.08$ & $62.11$\\
    &  {LLaVA1.5:MLP}& $7.80$ & $20.09$ & $27.60$ & $37.01$ & $51.27$ & $61.81$\\
    &  {Perceiver-Resampler}& $8.04$ &$21.58$ & $ 29.36$ &$ 39.23$ & $ 53.22$ & $63.97$\\
    &  {EM3} & $5.20$ & $ 13.48$ & $19.74$ & $28.13$ & $42.58$ & $55.63$\\
    &  {UniECS} & $6.60$ & $17.85$ & $25.07$ & $34.30$ & $48.95$ & $ 60.76$\\
    &  {NoteLLM-2} & \underline{8.19} & \underline{21.54} & \underline{29.83} & \underline{39.60} & \underline{54.36} & \underline{65.19}\\
    &  {TGQ-Former (Ours)} & \textbf{8.44} & \textbf{22.18} & \textbf{31.62} & \textbf{42.93} & \textbf{58.58} & \textbf{69.13}\\
    \midrule
    \multirow{7}{*}{\begin{tabular}{@{}c@{}}Chinese-CLIP-ViT-B/16 \\ + \\ Qwen3-1.7B\end{tabular}}&  {LLaVA:linear} & $7.77$ & $20.38$ & $27.99$ & $37.33$ & $51.17$ & $62.17$ \\
    &  {LLaVA1.5:MLP} & $7.68$ & $20.22$ & $28.02$ & $37.46$ & $51.63$ & $62.29$\\
    &  {Perceiver-Resampler}& 7.92 & \textbf{20.59} & \textbf{28.39} & 37.43 & 51.46 & 62.10\\
    &  {EM3} & $3.83$ & $11.06$ & $16.12$ & $23.35$ & $36.93$ & $49.87$\\
    &  {UniECS} & $6.84$ & $17.85$ & $24.87$ & $34.02$ & $48.49$ & $59.77$\\
    &  {NoteLLM-2}& 7.91 & 20.45 & 28.06 & \underline{37.73} & \underline{52.10} & \underline{62.92}\\
    &  {TGQ-Former (Ours)}& \textbf{8.13} & \underline{20.53} & \underline{28.34} & \textbf{38.05} & \textbf{52.36} & \textbf{63.20}\\
    \midrule
    \multirow{7}{*}{\begin{tabular}{@{}c@{}}Chinese-CLIP-ViT-B/16 \\ + \\ Baichuan2-7B-Chat\end{tabular}} &  {LLaVA:linear}& 7.53 & 19.64& 27.14  & 36.51  & 50.77 & 61.68 \\
    &  {LLaVA1.5:MLP}& 7.53  & 19.89& 27.41 & 36.55 & 50.93 & 61.91\\
    &  {Perceiver-Resampler} & 7.87 & 20.26 & 27.89 & 37.37 & 51.77 & 62.46\\
    &  {EM3} & 4.40 & 12.42 & 18.41& 26.80 & 41.48& 54.57 \\
    &  {UniECS}& 7.50  & 19.87& 27.84& 37.83 & 52.64 & 63.85\\
    &  {NoteLLM-2} & \underline{8.07}  & \underline{20.95} & \underline{29.16} & \underline{39.04} & \underline{53.55} & \underline{64.19}\\
    &  {TGQ-Former (Ours)}& \textbf{8.18} & \textbf{21.34} & \textbf{29.40} & \textbf{39.36} & \textbf{54.07}  & \textbf{64.57} \\
    \bottomrule
  \end{tabular}}
\end{table*}
\subsubsection{\textbf{Baselines.}}
\label{sec:baseline}
We compare with representative \emph{connector/fusion designs} that have been widely used to bridge frozen vision encoders and LLMs in MLLM/MLRM literature:
\begin{itemize}
    \item \textbf{LLaVA:Linear}~\cite{LLaVALinear}: a single linear projection that maps visual features into the LLM embedding space.
    \item \textbf{LLaVA-1.5:MLP}~\cite{LLaVAMLP}: a two-layer MLP for non-linear feature transformation.
    \item \textbf{Perceiver-Resampler} (Flamingo-style)~\cite{alayrac2022flamingo}: a perceiver-resampler module that compresses and aligns visual tokens with language inputs.
\end{itemize}

We further include recent multimodal retrieval/recommendation methods with similar MLRM-style pipelines:
\begin{itemize}
    \item \textbf{NoteLLM-2}~\cite{zhang2025notellm}: a BLIP-2 Q-Former based connector with design choices tailored for retrieval-style recommendation.
    \item \textbf{EM3}~\cite{deng2024end}: an FQ-Former that uses trainable queries and Transformer layers to fuse vision and text into fixed-length embeddings.
    \item \textbf{UniECS}~\cite{liang2025uniecs}: a gated cross-modal fusion mechanism originally proposed for e-commerce search; we adapt it to our I2I setting for comparison.
\end{itemize}

\subsubsection{\textbf{Metrics.}}
We evaluate I2I retrieval using Hit Rate (H@K)~\cite{tamm2021quality} at $K\in\{1,5,10,20,50,100\}$.
For each query item $i$ in the test set, we rank \emph{all} candidate items in the test item pool (approximately 134K unique items) by cosine similarity between their learned embeddings.
H@K is computed as whether at least one ground-truth positive item of $i$ appears in the top-$K$ ranked list.
All methods are evaluated under the same full-pool retrieval protocol.

\subsubsection{\textbf{Implementation Details.}}
We implement all models in PyTorch and train them with DDP on up to 8 H100 GPUs.
Unless otherwise noted, all methods use the same frozen Chinese-CLIP-ViT-B/16~\cite{yang2022chinese} visual encoder and the same Qwen3-Embedding-0.6B~\cite{zhang2025qwen3} backbone in the prompting pipeline.
We train with an InfoNCE objective and obtain a 256-dimensional item embedding via a linear projection from the LLM hidden state at the designated summary position.
Additional details are deferred to Appendix~\ref{app:impl}.



\subsection{Overall Performance Evaluation (RQ1)}

We evaluate the overall I2I retrieval performance of our framework against the baselines in Sec.~\ref{sec:baseline}.
All methods share the same vision encoder, prompting interface, and contrastive training objective; only the connector/fusion module differs.
We report results with three backbones in the MLRM prompting pipeline: Qwen3-Embedding-0.6B~\cite{zhang2025qwen3}, Qwen3-1.7B~\cite{yang2025qwen3}, and Baichuan2-7B-Chat~\cite{yang2023baichuan}, to assess generalizability across model families (details in Appendix Table~\ref{tab:llms}).
For Qwen3-1.7B and Baichuan2-7B-Chat, we follow common practice~\cite{zhang2025qwen3,zhang2025notellm} and use the hidden state at the designated summary position (e.g., the last token) as the item embedding.

Table~\ref{tab:overall_performance} shows that our method consistently outperforms all baselines across $K\in\{1,5,10,20,50,100\}$ under the full-pool retrieval protocol over the $\sim$134K test candidates.
With the Qwen3-Embedding-0.6B backbone, our model improves H@100 from 65.19 (NoteLLM-2) to 69.13, corresponding to +3.94 absolute points (6.04\% relative).
Compared with the Perceiver-Resampler connector, we achieve +5.16 absolute points at H@100 (8.07\% relative).
Improvements over linear/MLP connectors are also substantial, indicating that query-based tokenization with reliability-aware calibration is more effective for our noisy e-commerce imagery than directly projecting dense visual features.

Across backbones, we observe that the embedding-oriented Qwen3-Embedding-0.6B consistently yields strong performance, sometimes surpassing larger generative LLM backbones.
A plausible explanation is that embedding models are explicitly trained for representation learning and similarity-based retrieval, which better matches the objective of I2I embedding learning.

\subsection{Performance Comparison with MLLMs (RQ2)}
We compare our MLRM framework with end-to-end Multimodal LLMs (MLLMs) that directly encode images and text into a unified embedding.
We select representative 2--4B-class open-source MLLMs, including InternVL3-2B-Instruct~\cite{zhu2025internvl3}, Qwen2-VL-2B-Ins-truct~\cite{wang2024qwen2}, Qwen2.5-VL-3B-Instruct~\cite{bai2025qwen2} and Qwen3-VL-4B-Instruct~\cite{Bai2025Qwen3VLTR}.
For preprocessing, we follow each model's recommended configuration: images are resized to $224\times224$ by default, while InternVL3 uses its official $448\times448$ setting.
We report both retrieval quality (H@K) under the same full-pool evaluation protocol and the inference compute cost in GFLOPs per item embedding.

\begin{table}[tbh]
    \centering
    \setlength{\tabcolsep}{1.5pt} 
    \caption{The contrast between using MLLMs and using MLRMs with TGQ-Former.} 
    \resizebox{\linewidth}{!}{%
    \begin{tabular}{@{}lccccccc@{}}
        \toprule
        \textbf{Model}  & H@1 & H@5 & H@10 & H@20 & H@50 & H@100 & \textit{GFLOPs}\\ \midrule
        \multicolumn{8}{c}{\textbf{Using MLLMs}} \\ \midrule
        InternVL3-2B-Instruct   & 7.10  & 18.67 & 26.40 & 35.86 & 50.56 & 61.83 & \textit{724.71} \\
        Qwen2-VL-2B-Instruct    & 7.18  & 18.86 & 26.33 & 35.93 & 50.54 & 61.59 & \textit{298.91} \\
        Qwen2.5-VL-3B-Instruct  & \underline{8.48}  & \underline{22.07} & \underline{30.16} & 39.89 & 54.03 & 64.68 & \textit{450.16} \\
        Qwen3-VL-4B-Instruct  & 8.10 & 21.44 & 29.90 & \underline{39.98} & \underline{54.63} & \underline{65.68} & \textit{592.23} \\
        \midrule
        \multicolumn{8}{c}{\textbf{Using MLRMs with TGQ-Former}} \\ \midrule
        \begin{tabular}{@{}l@{}}Chinese-CLIP-ViT-B/16 + \\ Qwen3-Embedding-0.6B\end{tabular}   & 8.44 & 22.18 & 31.62 & 42.93 & 58.58 & 69.13 & \textit{35.43} \\
        \bottomrule
    \end{tabular}
    }
    \label{tab:mllm_comparison}
\end{table}

Table~\ref{tab:mllm_comparison} shows that our method achieves consistently better retrieval quality than all compared MLLMs while being substantially more compute-efficient.
Compared with the strongest MLLM baseline, Qwen3-VL-4B-Instruct, our model improves H@100 from 65.68 to 69.13 (+3.45 absolute points), and also yields gains at other $K$ values.
Meanwhile, our MLRM requires 8–20× fewer GFLOPs (35.43 vs 298--725 GFLOPs), highlighting the advantage of lightweight connectors and retrieval-oriented embedding backbones for high-throughput I2I retrieval.

\subsection{Ablation Study (RQ3)}
\label{sec:ablation}

We ablate key components of TGQ-Former under the same setting (Chinese-CLIP-ViT-B/16 + Qwen3-Embedding-0.6B) and full-pool retrieval over $\sim$134K test candidates. Results are reported in Table~\ref{tab:ablation}.

\textbf{Hybrid-Query Connector.}
Starting from a Q-Former baseline with exploratory queries only, replacing them with metadata-anchored semantic queries yields a clear gain, improving H@100 from 65.19 to 67.45.
Combining semantic and exploratory queries further improves performance, raising H@100 to 68.11, indicating that the exploratory stream provides complementary cues beyond metadata.

\textbf{Dual-Gated Vector Modulation.}
Adding Dual-Gated Vector Modulation on top of hybrid queries consistently boosts retrieval quality, improving H@100 from 68.11 to 68.90, supporting the need to calibrate the two streams separately under noisy poster-style listings.

\textbf{Redundancy-Reduction Regularizer.}
Finally, introducing the Redundancy-Reduction Regularizer brings additional improvement, increasing H@100 from 68.90 to 69.13.
Overall, the full model achieves the best results, demonstrating that hybrid queries, dual-gated calibration, and redundancy reduction contribute additively to robust I2I retrieval.

\begin{table*}[t]
\centering
\caption{Ablation study of our framework. All variants share the same vision encoder and LLM backbone; only the connector/modulation components differ.}
\label{tab:ablation}
\resizebox{0.92\linewidth}{!}{
\begin{tabular}{lccccccc}
\toprule
\textbf{Variant} &
\textbf{Metadata-anchored Queries} &
\textbf{Exploratory Queries} &
\textbf{Gating} &
$\boldsymbol{\mathcal{L}_{rr}}$ &
\textbf{H@20} & \textbf{H@50} & \textbf{H@100} \\
\midrule
(a) Q-Former(Exploratory-only) & -- & $\checkmark$ & -- & -- & 39.60 & 54.36 & 65.19 \\
(b) Metadata-anchored-only & $\checkmark$ & -- & -- & -- & 41.73 & 56.45 & 67.45 \\  
(c) Hybrid queries & $\checkmark$ & $\checkmark$ & -- & -- & 42.30 & 57.32 & 68.11 \\
(d) + Dual-gated Vector Modulation & $\checkmark$ & $\checkmark$ & $\checkmark$ & -- & 42.77 & 58.01 & 68.90 \\
(e) + Dual-Gate + RR (full) & $\checkmark$ & $\checkmark$ & $\checkmark$ & $\checkmark$ & \textbf{42.93} & \textbf{58.58} & \textbf{69.13} \\
\bottomrule
\end{tabular}
}
\end{table*}

\subsection{Qualitative Analysis of Disentangled Queries (RQ4)}
\label{sec:attn_vis}

\begin{figure}[!t]
\centering
\includegraphics[width=0.9\linewidth]{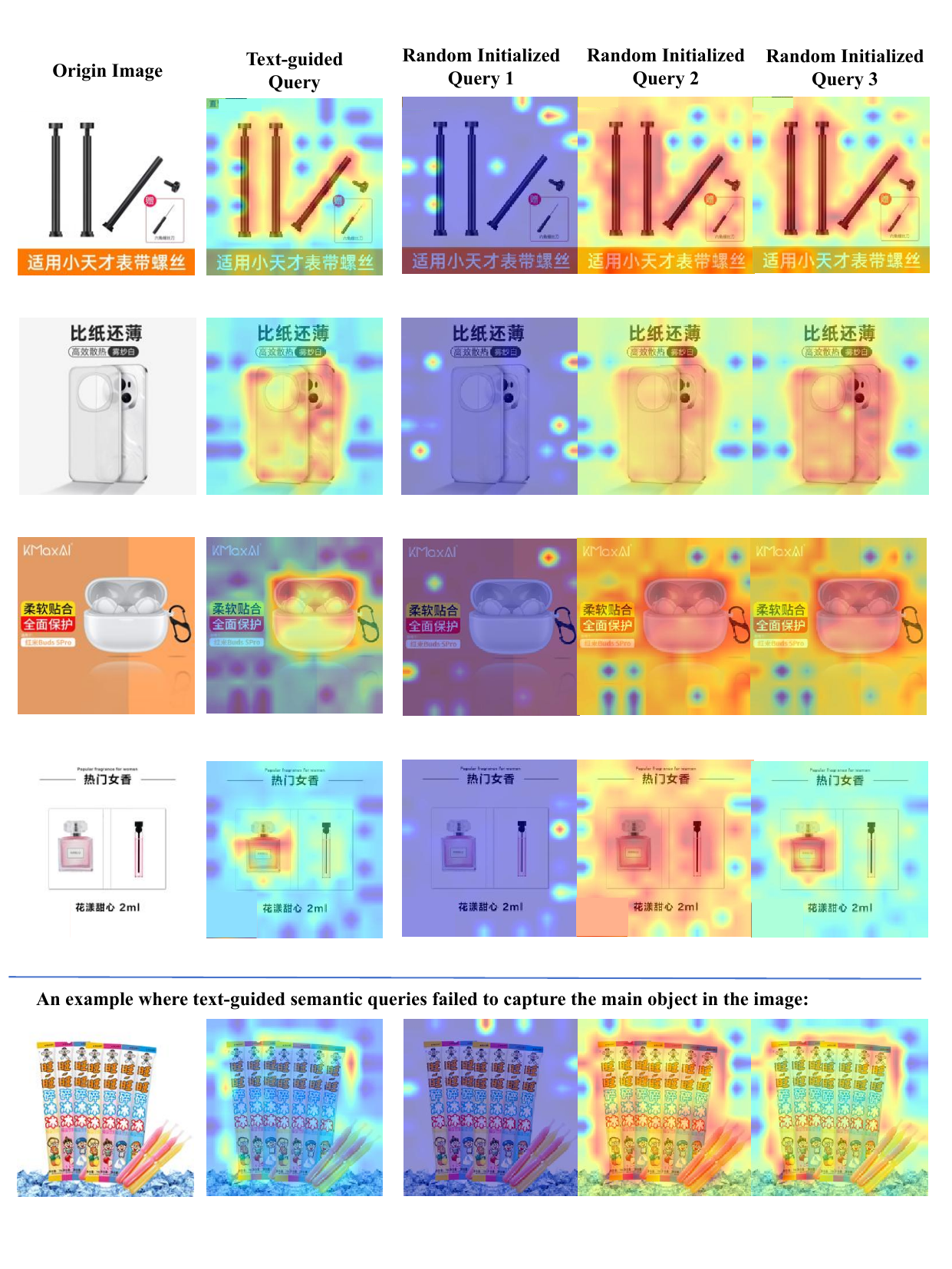}
\caption{Cross-attention visualization of different query types in TGQ-Former.
Each row corresponds to one product image.
The first column shows the original image.
The second column shows the averaged cross-attention map of metadata-anchored semantic queries (averaged over queries and heads).
The third to fifth columns show cross-attention maps of three representative exploratory queries, respectively.}
\label{pic_attn}
\end{figure}

To qualitatively examine how TGQ-Former disentangles metadata-anchored and exploratory visual evidence in poster-style listings, we visualize cross-attention maps on selected test examples.
Specifically, we extract the multi-head cross-attention weights from the last HQC layer and average them over attention heads.
Fig.~\ref{pic_attn} compares the attended regions of metadata-anchored semantic queries and exploratory queries.

As shown in Fig.~\ref{pic_attn}, semantic queries typically focus on product-relevant regions that are consistent with the metadata semantics, while largely avoiding visually salient but product-irrelevant overlay artifacts and decorative layouts.
In contrast, exploratory queries exhibit more diverse attention patterns, capturing both global context and fine-grained appearance cues that may not be explicitly specified by metadata.
These qualitative patterns are consistent with the intended roles of the two query streams.

We also observe complementary behavior in challenging cases.
When semantic queries attend to less informative regions in some cases, exploratory queries may still localize product-relevant evidence, providing complementary signals beyond metadata anchoring.

\begin{figure}[!t]
    \centering
    \includegraphics[width=0.7\linewidth]{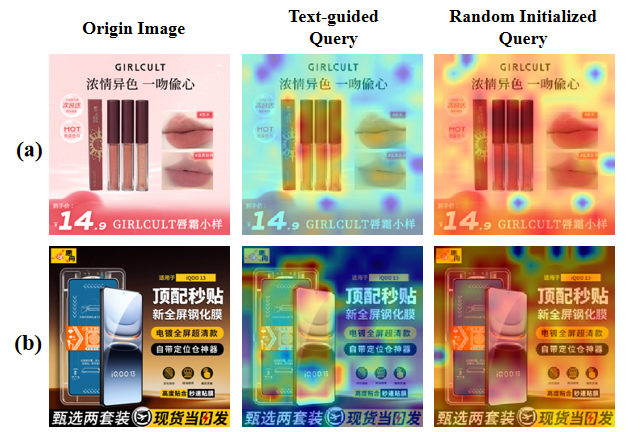}
    \caption{Qualitative examples from poster-style listings with dense overlay artifacts and clutter.
For each item, we show the original image (left), the averaged cross-attention map of metadata-anchored semantic queries (middle), and the averaged map of exploratory queries (right).}
    \label{fig:noisy_img}
\end{figure}

Fig.~\ref{fig:noisy_img} further illustrates examples with dense promotional overlays and clutter.
Even when high-contrast overlays occupy large portions of the image, semantic queries tend to localize product-relevant regions more consistently, while exploratory queries capture broader appearance patterns that can complement metadata-guided extraction.

\section{Conclusion}
This paper studied multimodal item-to-item (I2I) retrieval in e-commerce, where poster-style images often contain promotional overlays and clutter, introducing spurious cues and degrading fine-grained retrieval. We proposed TGQ-Former, a text-guided framework that performs \emph{adaptive disentanglement} by producing two complementary visual streams: metadata-anchored semantic queries capture metadata-consistent evidence, while exploratory queries capture complementary patterns beyond explicit metadata. To improve robustness under heterogeneous noise, we introduced Dual-Gated Vector Modulation to calibrate the streams using cross-modal agreement and image-derived cues, and a Redundancy-Reduction Regularizer to encourage complementarity. Experiments on a large-scale real-world dataset with full-pool retrieval show consistent gains over strong connector baselines and end-to-end MLLMs at substantially lower inference compute.

\bibliographystyle{ACM-Reference-Format}
\bibliography{sample-base}

\newpage

\appendix

\section{Model Details}

We present the details of the three LLMs employed in our method in Table~\ref{tab:llms}, and the details of the four MLLMs used for comparative experiments in Table~\ref{tab:MLLMs_detail}. Since our platform is designed for the Chinese market, all selected models provide support for Chinese processing.

\begin{table}[h]
  \caption{Model settings of LLMs}
  \label{tab:llms}
   \small
  \renewcommand{\arraystretch}{0.9}
  \setlength{\tabcolsep}{6pt}
   \resizebox{\linewidth}{!}{%
  \begin{tabular}{lccc}
    \toprule
    \textbf{Configuration}&\textbf{Qwen3-Embedding-0.6B}&\textbf{Qwen3-1.7B}&\textbf{Baichuan2-7B}\\
    \midrule
    \# LLM layers & 28 & 28 & 32\\
    \# LLM attention heads & 16 & 16 & 32\\
    Vocab size & 151669 & 151936 & 125696 \\
    LLM hidden size $d_{llm}$ & 1024 & 2048 & 4096\\
    LLM intermediate size & 3072 & 6144 & 4864\\
  \bottomrule
\end{tabular}}
\end{table}

\begin{table}[h]
  \caption{Model settings of MLLMs}
  \label{tab:MLLMs_detail}
   \small
   \renewcommand{\arraystretch}{0.8}
   \setlength{\tabcolsep}{0.8pt}
   \resizebox{\linewidth}{!}{%
  \begin{tabular}{lcccc}
    \toprule
    \textbf{Configuration}&\textbf{InternVL3-2B}&\textbf{Qwen2-VL-2B}&\textbf{Qwen2.5-VL-3B}&\textbf{Qwen3-VL-4B}\\
    \midrule
    Vision encoder $V_{\theta}$ & InternViT& ViT& ViT& ViT\\
    Connector $C_{\pi}$ & MLP & MLP & MLP & MLP (vision projector)\\
    LLM $LLM_{\mu}$ & Qwen2.5 & Qwen2& Qwen2.5& Qwen3\\
    \midrule
    \# Vision encoder layers & 24 & 32 & 32 & 24\\
    \# Vision encoder attention heads & 16 & 16 & 16 & 16\\
    Vision encoder hidden size $h_v$ & 1024 & 1536 & 1280 & 1024\\
    Vision encoder intermediate size & 4096 & 6144 & 3456 & 4096\\
    \midrule
    \# LLM layers & 28 & 28 & 36 & 36\\
    \# LLM attention heads & 12 & 12 & 16 & 32\\
    Vocab size & 151674 & 151936 & 151646 & 151936\\
    LLM hidden size $d_{llm}$ & 1536 & 1536 & 2048 & 2560\\
    LLM intermediate size & 8960 & 8960 & 4864 & 9728\\
  \bottomrule
\end{tabular}}
\end{table}

\section{Training Details and Hyperparameters}
\label{app:impl}

\begin{table}[t]
  \caption{Training hyperparameters for TGQ-Former.}
  \label{tab:hyperparams}
  \small
  \renewcommand{\arraystretch}{0.92}
  \setlength{\tabcolsep}{6pt}
  \centering
  \resizebox{\linewidth}{!}{%
  \begin{tabular}{lcc}
    \toprule
    \textbf{Category} & \textbf{Hyperparameter} & \textbf{Value / Setting} \\
    \midrule
    \multirow{4}{*}{\textit{General Settings}} 
      & Framework & PyTorch 2.6.0 \\
      & Language & Python 3.10 \\
      & GPUs & 8$\times$H100 (DDP) \\
      & Precision & fp16 \\
    \midrule
    \multirow{8}{*}{\textit{Optimization}} 
      & Optimizer & AdamW \\
      & Peak learning rate & $3\times10^{-5}$ \\
      & Weight decay & 1e-5 \\
      & $\beta_1, \beta_2$ & (0.9, 0.999) \\
      & $\epsilon$ & 1e-8 \\
      & Scheduler & Cosine annealing \\
      & Warm-up ratio & 10\% of total steps \\
      & Global batch size & 256 \\
    \midrule
    \multirow{6}{*}{\textit{LoRA Settings}} 
      & \textit{lora\_r} & 8 \\
      & \textit{lora\_alpha} & 16 \\
      & \textit{lora\_dropout} & 0 \\
      & Target modules & q,k,v,o projections \\
      & Applied layers & all Transformer blocks \\
      & Trainable params & LoRA only (LLM frozen otherwise) \\
    \midrule
    \multirow{5}{*}{\textit{TGQ-Former Structure}} 
      & \# random queries ($T_r$) & 3 \\
      & Query dim & 768 \\
      & Q-Former layers & 6 \\
      & Conv1D kernel/stride & 5/5 \\
      & Dual-gate MLP hidden dim & 1024 \\
    \midrule
    \multirow{4}{*}{\textit{Loss Settings}} 
      & Contrastive loss & InfoNCE \\
      & Temperature $\tau$ & 0.07 \\
      & $\lambda_{\text{rr}}$ (redundancy reduction) & 1 \\
      & Embedding normalization & L2-normalize before similarity \\
    \bottomrule
  \end{tabular}}
\end{table}

\paragraph{Backbones and preprocessing.}
We use a frozen CLIP image encoder and CLIP text encoder.
Specifically, we use Chinese-CLIP-ViT-B/16 with the standard CLIP preprocessing:
images are resized and center-cropped to $224\times224$ and normalized with the CLIP mean/std.
Item titles are tokenized with the CLIP tokenizer, truncated to 50 tokens.
The LLM backbone is Qwen3-Embedding-0.6B; we use its default tokenizer with a maximum length of 8192 tokens.
Unless stated otherwise, no additional image or text augmentation is applied.

\paragraph{Training.}
We train HQC and the dual-gating modules, and apply LoRA to the LLM backbone (only LoRA parameters are trainable in the LLM).
We use AdamW with the hyperparameters in Table~\ref{tab:hyperparams}.
Training is run with DDP on 8 H100 GPUs. The global batch size is 256, and the learning rate is $3 \times 10^{-5}$.
We train for 3 epochs with cosine annealing and warm up for 10\% of total steps.
We set the random seed to $\{42,43,44\}$ for Python/NumPy/PyTorch, and enable deterministic behavior where possible.

\paragraph{Evaluation.}
For evaluation, we perform full-pool retrieval over all test items ($\sim$134K candidates) for each query item.
We compute cosine similarity between the L2-normalized query embedding and each candidate embedding and rank all candidates.
We report Hit Rate H@K for $K\in\{ 1,5,10,20,50,100 \}$, where a query is considered a hit if any of its positives appears in the top-K ranked list.
If a query has multiple positives, we treat it as success when at least one positive is retrieved in top-K.

\section{Lightweight Cropping Preprocessing Baselines}
\label{app:crop_baselines}

A common heuristic to alleviate border-dominant promotional overlays (e.g., corner badges and banner-like decorations) is to crop the image before feeding it into the vision encoder. In this appendix, we evaluate simple, training-free preprocessing baselines that incur negligible overhead and do not rely on any learned detector.

\subsection{Setup}
\label{app:crop_setup}

\paragraph{Baselines.}
We compare three preprocessing variants:
\begin{itemize}
    \item \textbf{No-preprocessing}: the raw image is directly passed to the CLIP image processor.
    \item \textbf{Center-crop (0.9)}: center square crop retaining $0.9\times$ the shorter side, followed by the CLIP image processor.
    \item \textbf{Center-crop (0.8)}: center square crop retaining $0.8\times$ the shorter side, followed by the CLIP image processor.
\end{itemize}

\paragraph{Implementation details.}
Given an input image with width $W$ and height $H$, we compute $S=\min(W,H)$ and crop a centered square region with side length $s=\lfloor r\cdot S \rfloor$, where $r\in\{0.9,0.8\}$. The cropped image is then passed through the same CLIP image processor as in the main experiments (including its default resizing/normalization steps). All other components and hyperparameters are kept identical to the main setting.

\subsection{Results}
\label{app:crop_results}

Table~\ref{tab:crop_baselines} reports retrieval performance with lightweight center-cropping. While center-cropping slightly alleviates border-dominant overlays, the improvements remain limited and do not approach those of our method. This suggests that \emph{indiscriminate} suppression of non-product regions is insufficient in practice: useful contextual cues may co-exist with harmful overlay artifacts, calling for adaptive utilization of auxiliary signals rather than image-only preprocessing.

\begin{table}[t]
    \centering
    \small
    \setlength{\tabcolsep}{7pt}
    \begin{tabular}{lccc}
        \toprule
        \textbf{Preprocessing} & \textbf{H@20} & \textbf{H@50} & \textbf{H@100} \\
        \midrule
        No-preprocessing & 39.60 & 54.36 & 65.19 \\
        Center-crop (0.9) & 39.67 & 54.44 & 65.26 \\
        Center-crop (0.8) & 39.66 & 54.10 & 65.22 \\
        \bottomrule
    \end{tabular}
    \caption{Lightweight cropping preprocessing baselines. Center-cropping slightly alleviates border-dominant overlays but yields limited gains compared with our method.}
    \label{tab:crop_baselines}
\end{table}

\section{Open-Source Reproducibility on Amazon and Noise Robustness}
\label{sec:appendix_amazon_noise}

To mitigate reproducibility concerns under non-releasable proprietary data, we additionally evaluate our method on a public Amazon benchmark.
Notably, product images in this benchmark are typically clean and object-centric (often with plain backgrounds), which may under-represent the posterized and noisy conditions observed in real-world e-commerce.
We therefore further conduct a controlled robustness study by progressively injecting synthetic visual contamination into images, enabling a matched comparison under increasing noise levels.

\subsection{Dataset: Amazon Clothing, Shoes and Jewelry}
\label{sec:amazon_dataset}

We use the \textit{Amazon Clothing, Shoes and Jewelry} subset from the Amazon product metadata corpus (2018 version), and parse the metadata file \texttt{meta\_Clothing\_Shoes\_and\_Jewelry.json.gz}.
Each item is associated with (i) a primary product image URL stored in the \texttt{imUrl} field, and (ii) textual metadata including \texttt{title}, \texttt{brand}, and a hierarchical \texttt{categories} field.
Following common I2I retrieval setups, we construct positive pairs $(i,j)$ using dataset-provided \texttt{related} fields, including \texttt{also\_bought}, \texttt{bought\_together}, and \texttt{also\_viewed}.
Unless otherwise stated, we treat the related graph as directed and do not symmetrize edges.

\paragraph{Training set construction.}
We first build a valid item pool by scanning \texttt{meta\_Clothing\_Shoes\_and\_Jewelry.json.gz} and retaining items that satisfy:
(i) non-empty \texttt{title};
(ii) a non-empty \texttt{imUrl};
(iii) a valid category path with at least one category list, where we take the first category path and construct a 3-level hierarchy by truncation, and if its depth is less than 3 we pad by repeating the last level (``relaxed'' 3-level parsing);
and (iv) the item ASIN is \emph{not} in a pre-generated test ASIN blocklist (constructed below), ensuring train/test separation at the ASIN level.

For each retained item $i$, we enumerate its related targets $j$ from \texttt{also\_bought}, \texttt{bought\_together}, and \texttt{also\_viewed}, and keep a training pair $(i,j)$ only if $j$ is also in the valid item pool (and thus also not blocklisted).
We then cap the number of training pairs to 1{,}200{,}000 by shuffling with a fixed random seed (42) and taking the first 1{,}200{,}000 pairs as our training set.

\paragraph{Test set construction and candidate pool.}
We build the test set via pair-based random sampling over the same valid item pool used for training (items with non-empty title, a parsed 3-level category path, and an accessible image).
We first collect all valid ASINs, shuffle them with a fixed random seed (42), and traverse the shuffled list.
For each candidate query item $i$, we select a single ground-truth neighbor $j$ by scanning its \texttt{related} field and taking the first valid target that also lies in the valid pool.
If such a target exists, we add the pair $(i,j)$ to the test pairs and include both $i$ and $j$ in the test ASIN set.
We repeat this process until obtaining 60K test pairs.
Items without any valid related target are skipped.

We define the evaluation candidate pool as the set of all unique ASINs appearing in the sampled test pairs (i.e., all query items and their selected targets).
For evaluation, each query is associated with a \emph{single} ground-truth target (Top-1) constructed as above, and retrieval ranks all items in the candidate pool.
Finally, to avoid ASIN-level overlap between train and test, we export all test ASINs as a blocklist and exclude them from training set construction.

\subsection{Progressive Noise Injection}
\label{sec:amazon_noise}

To emulate real-world e-commerce posterization and background clutter, we generate four image conditions with increasing severity:
\textbf{(i) Clean}, \textbf{(ii) Light}, \textbf{(iii) Medium}, and \textbf{(iv) Heavy}.
For reproducibility, we use a fixed random seed (42) and generate one deterministic corrupted view per test image for each severity level.

\paragraph{Noise operators.}
We consider two synthetic corruption operators:

\textbf{Background replacement.}
We estimate a foreground mask by thresholding near-white pixels in RGB: a pixel is treated as background if $\min(R,G,B) \ge 240$.
We keep the complement as foreground and composite it onto a randomly sampled background crop from another image.
The background crop is uniformly sampled in location and resized to the original image size; we optionally apply a Gaussian blur with radius $r_{\text{blur}}=2$ to the background before compositing.

\textbf{Semantic overlay noise.}
We add poster-like overlay artifacts that mimic common e-commerce decorations, including (i) \emph{burst badges}, (ii) \emph{corner banners}, and (iii) \emph{top/bottom bars}.
For each image, with probability $p(\text{overlay})$ we sample up to N overlays uniformly from the three types and render it near the image border.
Concretely, the overlay anchor is sampled from one of the four corners with a 10\% margin: the top-left/top-right/bottom-left/bottom-right region, where the $(x,y)$ position is uniformly sampled within the corresponding margin box.
The overlay geometry is scaled relative to image size: burst badges use a star-shaped polygon with radius in $[0.15,0.20]\cdot\min(W,H)$; corner banners are filled triangular patches with side length in $[0.35,0.45]\cdot\min(W,H)$ (top-left or top-right); and bars are rectangles with height in $[0.15,0.20]\cdot H$ placed at the top or bottom.
We render short marketing text in white on top of the overlay using a fixed font (Arial when available, otherwise a default font), with text sampled from small, fixed vocabularies (e.g., badge: \{\texttt{HOT}, \texttt{SALE}, \texttt{50\%}, \texttt{NEW}, \texttt{No.1}, \texttt{TOP}, \texttt{9.9}, \texttt{Best}\}; banner/bar: \{\texttt{Free Shipping}, \texttt{Flash Sale}, \texttt{New Arrival}, \texttt{Limited Offer}\}).

\paragraph{Severity levels.}
We instantiate four severities by controlling the application probability and intensity of the above operators (Table~\ref{tab:noise_hparams}).
When both operators are enabled, we apply background replacement first and then overlay noise.

\begin{table}[t]
    \centering
    \small
    \caption{Noise hyperparameters for the progressive robustness test.}
    \label{tab:noise_hparams}
    \setlength{\tabcolsep}{4pt}
    \begin{tabular}{lccc}
        \toprule
        \textbf{Severity} & $p(\text{bg repl.})$ & $p(\text{overlay})$ & \#overlays \\
        \midrule
        Clean  & 0.0 & 0.0 & 0 \\
        Light  & 0.0 & 0.4 & 1 \\
        Medium & 0.5 & 0.7 & 3 \\
        Heavy  & 0.8 & 0.9 & 5 \\
        \bottomrule
    \end{tabular}
\end{table}

\subsection{Compared Methods and Evaluation Protocol}
\label{sec:amazon_protocol}

We compare our method with the strongest connector-based baseline:
\textbf{NoteLLM-2}~\cite{zhang2025notellm}.
All methods share the same vision encoder and LLM embedding backbone as in the main experiments unless otherwise stated.
We follow the standard top-$K$ retrieval evaluation and report Hit Rate@K (H@K) for $K \in \{10,20,50\}$.
For each test query, candidates are all items in the candidate pool (all unique ASINs in the sampled test pairs), and the ground-truth neighbor is the paired target in the test pairs.

\subsection{Results: Accuracy under Increasing Noise}
\label{sec:amazon_results}

Table~\ref{tab:amazon_noise_results} reports retrieval accuracy on Amazon under progressive noise.
We highlight two aspects:
(i) \textbf{clean-setting performance}, which reflects transferability to object-centric images;
(ii) \textbf{degradation curves} as noise increases, which measures robustness to posterization-like corruptions.
We expect both methods to perform similarly on Clean images, while our method should degrade more gracefully under Medium/Heavy noise due to its reliability-aware visual extraction and calibration.

\begin{table}[t]
    \centering
    \small
    \caption{I2I retrieval on Amazon Clothing, Shoes and Jewelry under progressive noise.}
    \label{tab:amazon_noise_results}
    \setlength{\tabcolsep}{4pt}
    \begin{tabular}{llcccc}
        \toprule
        \textbf{Setting} & \textbf{Method} & H@10 & H@20 & H@50 \\
        \midrule
        \multirow{2}{*}{Clean}
            & NoteLLM-2 & 46.05 & 57.12 & 69.41 \\
            & Ours      & 46.02 & 57.18 & 69.46 \\
        \midrule
        \multirow{2}{*}{Light}
            & NoteLLM-2 & 45.79 & 56.76 & 68.93 \\
            & Ours      & 45.92 & 57.03 & 69.27 \\
        \midrule
        \multirow{2}{*}{Medium}
            & NoteLLM-2 & 45.01 & 56.05 & 68.11 \\
            & Ours      & 45.33 & 56.48 & 68.73 \\
        \midrule
        \multirow{2}{*}{Heavy}
            & NoteLLM-2 & 44.28 & 55.37 & 67.55 \\
            & Ours      & 44.91 & 56.04 & 68.33 \\
        \bottomrule
    \end{tabular}
\end{table}

\paragraph{Discussion.}
On the clean benchmark, both approaches operate on high-quality object-centric images, and the gap can be small.
As we increase the noise severity, the baseline can be distracted by high-contrast overlays and altered backgrounds, leading to larger performance drops.
In contrast, our method is designed to selectively trust visual evidence conditioned on its reliability, and thus should maintain more stable retrieval quality under Medium/Heavy noise.

\end{document}